\documentclass[twocolumn,aps,floatfix,superscriptaddress]{revtex4}
\usepackage{graphicx,amssymb,amsfonts,amsmath,color,bm,hyperref,subfig}
\usepackage{xcolor}
\hypersetup{
    colorlinks,
    linkcolor={blue},
    citecolor={blue},
    urlcolor={black}
}
\newcommand{\rmd}{\textrm{d}}
\newcommand{\rme}{\textrm{e}}
\newcommand{\rmi}{\textrm{i}}

\begin{document}
\title{ Work distribution function for a Brownian particle driven by a nonconservative force}
\author{Bappa Saha}
\email[]{bsaha@iitk.ac.in}
\affiliation{Department of Physics, Indian Institute of Technology,
Kanpur-208 016, India}
\author{Sutapa Mukherji}
\email[]{sutapam@iitk.ac.in}
\email[(On leave from Department  of Physics, Indian Institute of Technology, Kanpur- 208 016, India)]{}
\affiliation{Department of Protein Chemistry and Technology, 
Central Food Technological Research Institute, 
Mysore-570 020, India}
\date{\today}
\begin{abstract}
We  derive the distribution function of work performed by a harmonic force acting on a uniformly dragged Brownian particle subjected to a rotational torque. Following the  Onsager and 
Machlup's functional integral approach, we obtain 
the transition probability of finding the 
Brownian particle at a particular  position at time 
$t$ given 
that it started the journey from a specific 
location at an earlier time. 
The difference between 
the forward and the time-reversed form of the 
generalized Onsager-Machlup's Lagrangian is 
identified as the rate of medium entropy production 
which further helps us develop the stochastic 
thermodynamics formalism for our model.
The probability distribution for the 
work done by the harmonic trap is evaluated 
for an equilibrium initial condition. 
Although this distribution has a Gaussian form, 
it is found that  the  distribution does not 
satisfy the conventional work fluctuation theorem.    
\end{abstract}
\maketitle

\section{Introduction}
Fluctuations are ubiquitous in most of the physical processes we observe in nature. In general, though the effect of 
 fluctuations could be ignored for 
the coarse grained description of the 
average thermodynamic quantities, these are
no longer negligible for  small systems, 
like, biomolecules, nano-machines etc. 
For instance, the movement of a molecular motor 
of nanometer size is affected 
significantly
 by the mean thermal energy of the surrounding 
medium which causes the motor's position to 
fluctuate.  In linear irreversible processes, 
these fluctuations obey a general principle 
known as the 
fluctuation-dissipation theorem 
\cite{kubo,kubo_book}. 
However, in the case of  
systems far away from equilibrium, 
the assumptions of the 
linear-response theory are not valid 
since the fluctuations of the 
thermodynamic variables do not follow the 
 linear law of relaxation.

Since the finding of the fluctuation relations for the  probabilities of
observing the entropy producing and consuming trajectories in  simulations of shearing fluid nearly two decades ago \cite{evans1},  there has been a vast number of studies \cite{evans2,gallavotti,kurchan,jarzynskirel,
crooks,lebowitz,zon,zon_cohen1,zon_cohen2,harris, seifert2,seifert,seifert1,esposito,noneql_book,derrida}  
 extending the ideas and formalisms of linear irreversible thermodynamics to nonequilibrium systems. 
The emphasis of the fluctuation relations 
has been on quantifying the asymmetry between 
the probability distributions of a 
time-integrated quantity 
$\Sigma_t$ to have values $\pm W$ 
in the forward and time-reversed processes, 
and it is  expressed
 in the mathematical form as
$\frac{P_F(\Sigma_t=+W)}{P_R(\Sigma_{t}^{\dagger}=-W)}
=\exp(W/k_BT)$. $\Sigma_t$ 
could represent the work done 
by an external force, 
dissipated heat, 
entropy production etc, 
and $\Sigma_{t}^{\dagger}$ stands for the 
same thermodynamic quantity in the time-reversed 
process. These relations, valid for 
systems far-from-equilibrium, 
could be reduced to the 
fluctuation-dissipation theorem for 
near-equilibrium processes \cite{derrida}.  
Of the many ramifications of the 
fluctuation relations, the most notable results are 
the Jarzynski equality \cite{jarzynskirel} and 
Crooks work relation \cite{crooks}  
which relate the asymmetry to 
the free energy difference between the 
initial and final  equilibrium 
states which are connected through a
reversible or irreversible transformation.
Although, to prove the fluctuation  
relations, one does not always 
require to obtain the detailed distribution 
functions, to verify them experimentally, 
however, it is preferable to get 
the histograms of the relevant thermodynamic 
quantities.  In many situations 
\cite{ritort}, a detailed 
knowledge of the distribution of the 
nonequilibrium work might be 
essential to compute the equilibrium 
free energy difference utilizing 
the Jarzynski equality. Despite the 
significant progress in the theoretical 
studies of these distribution 
functions for work, heat, 
entropy production 
etc for a variety of nonequilibrium 
systems  \cite{zon,zon_cohen1,zon_cohen2,mazonka,
seifert3,imparato3,farago,ritort,imparato1,
park1,park2,sanjib1,sanjib2} 
(and the references therein), 
the  derivation of  
the exact analytical expressions for these 
distributions are not so many. 
This is   possibly due to the
 increasing difficulty in obtaining 
the analytical solutions in closed forms 
as the dynamics of the system becomes 
more complex. 

In this paper, we derive the 
  positional  probability 
distribution  
and  the work distribution function
for a Brownian particle 
using the Onsager-Machlup's functional 
integral approach. In their theory 
of linear relaxation processes, 
 Onsager and Machlup developed a  path integral approach \cite{onsager1,onsager2}  for the 
probability distribution functions 
of  trajectories in the configuration space 
with   specific initial and final states. 
Using a variational technique they subsequently
 showed that this approach leads to Onsager's minimal energy 
dissipation. 
Recently, there have been many efforts to 
generalize the Onsager-Machlup 
theory for systems in  nonequilibrium steady states \cite{bertini1,bertini2,bertini3,imparato2,taniguchi,cohen,maes,jstat2}. Extending this approach, Taniguchi and Cohen 
\cite{taniguchi,cohen}  obtained 
the distribution functions of work, dissipated heat 
for a uniformly dragged Brownian particle. 
Further,  after expressing the Onsager-Machlup Lagrangian 
in terms of the dissipation functions and the 
entropy production rate,  they showed how these 
fluctuating quantities satisfy the 
energy conservation law and the second law 
of thermodynamics. 
Similar variational scheme  has also 
been considered to obtain the 
probability distributions of thermodynamic 
quantities  for a variety of systems 
\cite{engel}.

In the present work, we 
consider  two dimensional motion of a 
Brownian particle which is subjected to a 
nonconservative force  and a confining harmonic 
potential with its  minimum moving  
with a uniform velocity $v$. 
The nonconservative force 
\begin{eqnarray}
f=-y n \hat {x}+x n \hat{y}, \label{noncons-force}
\end{eqnarray}
with  $n$ denoting 
the strength of the  force,
produces an anticlockwise drift about 
 the origin. 
The  nonconservative force 
can arise due to a  
vortex centered   at the origin of the 
system in the surrounding liquid \cite{vladimir}. 
Both the time dependent confining potential 
and the nonconservative force are 
responsible for breaking the detailed balance 
and driving the system out of equilibrium. 
For this model,  
we derive the joint  probability  
distribution which gives the probability 
of finding a Brownian particle at a given position
at a specific time. 
After defining the forward 
and the  time-reversed trajectories, we 
obtain the expression for the dissipated 
heat and the energy conservation relation. 
In the process, we identify various works, 
namely, the
work done by the harmonic trap
which can  be identified as the 
thermodynamic work \cite{jarzynskirel} 
and the  
mechanical work done by the nonconservative force 
\cite{park1,park2}. 
Finally, using a variational approach, 
we obtain the  distribution 
function for the work done by the trap. 
The work distribution has a Gaussian form 
with the mean and the variance  
 having  a nontrivial
 oscillatory dependence originating  
from the nonconservative force. Although, 
this distribution is Gaussian, it does not 
satisfy the conventional fluctuation relation.
Physically, the presence of the 
nonconservative force makes the role of the 
thermal fluctuation negligible. 
Similar deviation can also be found in \cite{ciliberto2,joubaud} for the motion of a torsion pendulum driven by an external time dependent torque.

The rest of the paper is organized in the following way. In section \ref{sec:themodel}, we introduce two 
coupled Langevin equations describing the motion 
of the Brownian particle in the $x-y$ plane. 
In section \ref{sec:probdist}, 
 we find out 
the joint probability  distribution for the position 
of the Brownian particle. Starting with the Fokker-Planck equation for the 
joint probability distribution, we express 
first the transition probability
 in a functional 
integral form  and subsequently 
evaluate the path integrals 
employing Onsager-Machlup's 
variational approach. The  
joint probability distribution 
is obtained from the expression of 
the transition probability. From the generalized Onsager-Machlup  Lagrangian, 
we define in  section \ref{sec:fewrelations} 
the thermodynamic work, mechanical work, 
dissipated heat and the entropy production of the 
medium. We further show here the validity of 
the energy conservation  relation 
and the second law of thermodynamics 
 for a trajectory. 
In section \ref{sec:workdist1}, we 
obtain the explicit form of the distribution 
function of the thermodynamic work.  We 
summarize our findings and discuss 
possible future research on this model 
in section \ref{sec:summary}. 
Some of the details of the derivations 
and algebra are presented in the appendices.

\section{The model}\label{sec:themodel}
The Brownian   particle considered here  
moves in a plane under the influence 
of the potential 
$U[x(t),y(t),t]=\frac{k}{2}\left\lbrace(x(t)-vt)^2+y(t)^2\right\rbrace$ and a  nonconservative force 
as given in equation (\ref{noncons-force}) 
\cite{vladimir,park2}.  
 Here $v$ is the constant velocity
 with which the harmonic trap moves 
 along the $x$ direction. 

The overdamped Langevin equations describing the motion of the particle in the $x$-$y$ plane are expressed as
\begin{eqnarray}
\dot{x}&=&-\frac{1}{\tau_r}(x-vt)- \sigma y +\frac{1}{\alpha} \xi_x(t) ,\\
\dot{y}&=&-\frac{1}{\tau_r}y+\sigma x +\frac{1}{\alpha}\xi_y(t) ,
\end{eqnarray}
where, $\alpha$ is the viscous drag coefficient, $\sigma=n/\alpha$ and $\tau_r=\alpha/k$ is the relaxation time of the trap. The overdots imply derivatives with respect to time, $t$.
$\xi_x(t)$ and $\xi_y(t)$ represent the Gaussian-distributed thermal noise  having   zero mean and correlations as  
\begin{eqnarray}
\left\langle \xi_i(t)\xi_j(t^\prime)\right\rangle = \frac{2\alpha}{\beta}\ \delta_{ij}\delta(t-t^\prime).
\end{eqnarray}
Here $\beta=(k_BT)^{-1}$ and $i,\ j$ correspond to the 
$x, \ y$ coordinates.

\section{Probability distribution functions related to the location of the particle}\label{sec:probdist}
The joint probability distribution is defined as the 
probability of finding the Brownian particle at position 
$(x,y)$ at time $t$.
The time evolution of the joint  probability
 distribution is governed by the Fokker-Planck 
equation which can be written as
\begin{eqnarray}
\frac{\partial \rho(x,y,t)}{\partial t}=\hat{L}_{\rm FP}\ \rho(x,y,t), \label{fpeqn}
\end{eqnarray}
where, the Fokker-Planck operator $\hat{L}_{\rm FP}$ is given by,
\begin{eqnarray}
\hat{L}_{\rm FP}&=&\frac{\partial}{\partial x}\left\{(x-vt)/\tau_r +\sigma y\right\}-\frac{\partial}{\partial y}\left\{-y/\tau_r +\sigma x\right\}\nonumber\\
 &&+D\frac{\partial^2}{\partial x^2} +D\frac{\partial^2}{\partial y^2}, \label{lfp}
\end{eqnarray}
with the diffusion constant, $D=1/(\alpha \beta)$. 
Let us consider the simplest situation where 
  $v=\sigma=0$. 
Writing the Fokker-Planck equation in terms of the 
probability current, ${\bf J}_\rho$, we have  
\begin{eqnarray}
\frac{\partial\rho}{\partial t}=-\nabla\cdot {\bf J}_\rho,
\end{eqnarray}
where 
\begin{eqnarray}{\bf J}_\rho={\hat{x}} \left[\frac{x\rho}{\tau_r}+
\frac{D\partial\rho}{\partial x}\right]+
{\hat{y}}\left[\frac{y \rho}{\tau_r}+
\frac{D\partial\rho}{\partial y}\right].
\end{eqnarray}
Substituting $\rho=\exp[-\beta U]$ (with $v=0$), we obtain a vanishing stationary current corresponding to the equilibrium
 situation of a fixed 
harmonic potential. This is the situation where the 
detailed balance is satisfied. 
The nonconservative force which cannot be expressed 
in the form of a gradient of a potential violates the detailed 
balance.  

 $\rho(x,y,t)$ can be 
expressed in terms of the transition probability $P(x,y,t \mid x_0,y_0,t_0)$ which gives the 
probability of finding a Brownian particle at position 
$(x,y)$ at time $t$,
 given that it was at $(x_0,y_0)$ at an initial 
time $t_0$. For simplicity, we 
choose $t_0=0$ and denote the 
transition probability as $P(x,y,t\mid x_0,y_0)$.
The Fokker-Planck equation (\ref{fpeqn}) allows us 
to obtain  the following path integral  
representation for the 
transition probability \cite{risken} (some of the details of the derivation are presented in appendix-\ref{app:appendixA})
\begin{flalign}
 P(x_f,y_f,t_f \mid x_0,y_0)=&\int {\cal D}[x(t)]\ \int {\cal D}[y(t)]&\nonumber\\
  &\times \exp\left(-\int_0^{t_f} {\cal L}(x,y,\dot{x},\dot{y},t) \,\rmd t \right).& \label{functional_int}
\end{flalign} 
${\cal L}(x,y,\dot{x},\dot{y},t)$ is the generalized Onsager-Machlup Lagrangian having the form
\begin{eqnarray}
{\cal L}(x,y,\dot{x},\dot{y},t)&=&\frac{1}{4D}\bigl[(\dot{x}+(x-vt)/\tau_r +\sigma y)^2\nonumber\\
&&+(\dot{y}+y/\tau_r -\sigma x)^2\bigr] .\label{om_lagrangian}
\end{eqnarray}

We evaluate the functional integrals 
 (\ref{functional_int}) using a variational 
approach \cite{onsager1,onsager2,wiegel,taniguchi,cohen}, 
in which  one evaluates 
the most probable path that gives the most significant 
contribution  to the functional integral. This is expected 
to be a good approximation for low noise strength or 
$D\rightarrow 0$ \cite{wiegel}.  
The approximate form of the  transition 
probability can be written  as
\begin{flalign}
&P(x_f,y_f,t_f \mid x_0,y_0) \propto \exp\Bigl(-\Bigl[\int_0^{t_f} {\cal L}(x,y,\dot{x},\dot{y},t) \,\rmd t\Bigr]_{\rm min} \Bigr),& 
\label{functional_int1}
\end{flalign}
where the object in the square bracket is evaluated over the 
most probable path leading  to the largest value of the 
exponential. 
The most-probable path is determined by using the extremization  condition
\begin{equation}
\delta \int_0^{t_f} {\cal L}(x,y,\dot{x},\dot{y},t)\, \rmd t = 0 . \label{extremal}
\end{equation}
Doing a Taylor-expansion around the optimal values,  
$(x_c, y_c)$, and  
retaining fluctuations about the optimal path, 
$u=(x-x_c)$ and $z=(y-y_c)$ 
up to second order the transition probability in (\ref{functional_int}) can be written as,
\begin{flalign}
P(x_f,y_f,t_f \mid x_0,y_0)&=\phi(t_f)& \nonumber \\ 
&\times \exp\left(-\int_0^{t_f} {\cal L}(x_c,y_c,\dot{x}_c,\dot{y}_c,t) \,\rmd t \right).& \label{transition1}
\end{flalign}
Since we consider trajectories with fixed end-points 
and allow no variation at the end points, $u$ and $z$ 
satisfy Dirichlet boundary conditions 
$u(0)=0,\ u(t_f)=0, \ z(0)=0,\ z(t_f)=0$.
As a consequence of this, the  
fluctuation factor  
\begin{flalign}
&\phi(t_f)=\int_{u(0)=0}^{u(t_f)=0} {\cal D}[u(t)]\, \int_{z(0)=0}^{z(t_f)=0} {\cal D}[z(t)]&\nonumber\\
&\times \exp\biggl\{-\frac{1}{4D}\left(\left[\dot{u}+u/\tau_r+\sigma z\right]^2+\left[\dot{z}+z/\tau_r -\sigma u\right]^2\right)\biggr\}, &\label{fluctuationfactor}
\end{flalign}
is a function of the final time $t_f$ only.
Instead of evaluating  the path integrals in Eq. 
(\ref{fluctuationfactor}) directly, we determine  
$\phi(t_f)$ using the  normalization condition 
of the transition probability, i.e., the total probability of finding the Brownian particle  
at any final value of $(x_f,y_f)$ after time $t_f$ is unity. 
It can be proved, \textit{a posteriori}, from the normalization property of $P(x_f,y_f,t_f\mid x_0,y_0)\mid_{v=0}$, that the final expression of $\phi(t_f)$ has the form
\begin{equation}
\phi(t_f)=\frac{\left[1+\coth(t_f/\tau_r)\right]}{4\pi D\tau_r} \label{fluctuationfactor_final}.
\end{equation}

The extremization condition (\ref{extremal}) 
leads to two coupled Euler-Lagrange equations for the optimal paths $x_c(t)$ and $y_c(t)$ 
\begin{eqnarray}
\ddot{x}_c-(\sigma^2+1/\tau_r^2)x_c +2\sigma \dot{y}_c+v\left(\frac{t}{\tau_r^2}-\frac{1}{\tau_r}\right)=0\\ 
\ddot{y}_c-(\sigma^2+1/\tau_r^2)y_c - 2\sigma \dot{x}_c +v\sigma t/\tau_r =0.
\end{eqnarray}
The solutions of the above equations are 
\begin{eqnarray}
x_c(t)&=& v[bt-\tau_r +\sigma^2\tau_r^3]/b^2 \nonumber\\
&&+ \cosh(t/\tau_r)[a_2 \cos(\sigma t)-a_1 \sin(\sigma t)]\nonumber\\
&&+\sinh(t/\tau_r)[a_4 \cos(\sigma t)- a_3 \sin(\sigma t)]\label{optimal_x} ,\\ \nonumber
\\
y_c(t) &=& \sigma \tau_r v [bt-2\tau_r]/b^2 \nonumber \\&&+ \cosh(t/\tau_r)[a_1 \cos(\sigma t)+a_2\sin(\sigma t)]
\nonumber\\ &&+ \sinh(t/\tau_r)[a_3\cos(\sigma t)+a_4\sin(\sigma t)]\label{optimal_y} ,
\end{eqnarray}
where $b=(1+\sigma^2\tau_r^2)$. The constants $a_1,\ a_2, \ a_3$ and $a_4$ are calculated using the boundary conditions $x_c(0)=\nobreak x_0,\ x_c(t_f)=x_f,\ y_c(0)=y_0,\ y_c(t_f)=y_f$. 
These constants are
\begin{eqnarray}
a_1&=&y_0+2\sigma v \tau_r^2 /b^2 ,\\
 a_2&=&x_0-v(\sigma^2\tau_r^3 -\tau_r)/b^2,\\
a_3&=&-\frac{{\rm csch}(t_f/\tau_r)}{b^2} \biggl\{ \cos(\sigma t_f) \left(-2\sigma \tau_r^2 v+b t_f v\sigma\tau_r-b^2 y_f\right)\nonumber\\ &&+(2\sigma\tau_r^2 v+b^2 y_0)\cosh(t_f/\tau_r)\nonumber\\ &&+\sin(\sigma t_f)\left(\tau_r v-\sigma^2\tau_r^3 v-b t_f v+b^2 x_f\right) \biggr\}, \\
a_4&=&\frac{{\rm csch}(t_f/\tau_r)}{b^2} \biggl\{ \cos(\sigma t_f)\left(\tau_r v-\sigma^2\tau_r^3 v-b t_f v+b^2 x_f \right)\nonumber\\ &&+\cosh(t_f/\tau_r)\left(-\tau_r v+\sigma^2\tau_r^3 v-b^2 x_0\right)\nonumber\\
&&+\sin(\sigma t_f)\left(2\sigma\tau_r^2 v-\sigma\tau_r b t_f v+b^2 y_f\right)\biggr\}\label{constants}.
\end{eqnarray}
Substituting Eqs.(\ref{fluctuationfactor}), (\ref{optimal_x}) and (\ref{optimal_y}) into Eq.(\ref{transition1}), and performing the integration over time, we have the complete expressions for $P(x_f,y_f,t_f\mid x_0,y_0)$ as
\begin{eqnarray}
P(x_f,y_f,t_f\mid x_0,y_0)=\frac{[1+\coth(t_f/\tau_r)]}{4\pi D\tau_r}\nonumber \\
\times  \exp \biggl\{ -\frac{(-1+\rme ^{2t_f/\tau_r})}{8D\tau_r} \left[(a_1+a_3)^2+(a_2+a_4)^2\right]\biggr\}. \label{transition_prob}
\end{eqnarray}

\begin{figure}[ht!]
  \centering
   \includegraphics[height=.38\textwidth]{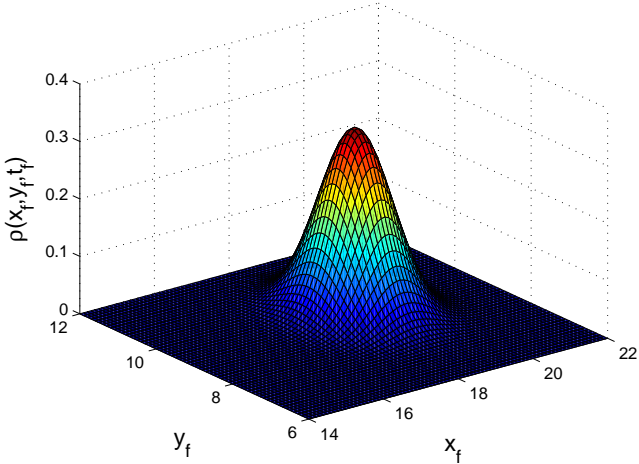}  
\caption{Plot of the joint distribution function $\rho(x_f,y_f,t_f)$ for the parameter values $D=0.05,\  \beta=10,$ $\tau_r=10,\ k=0.2,$ and $v=2$ at  final time $t_f=20$.}
\label{fig:samplefig1}
\end{figure}

To obtain the joint probability distribution $\rho(x_f,y_f,t_f)$, we integrate the above expression over the initial points, assuming that the system was initially in thermal equilibrium with the joint probability distribution \\
$\rho(x_0,y_0)=\frac{\beta k}{2\pi}\ \exp\left(-\frac{\beta k}{2}[x_0^2+y_0^2]\right)$.  So,
\begin{eqnarray}
\rho(x_f,y_f,t_f)&=&\int \rmd x_0\int \rmd y_0 \rho(x_0,y_0) P(x_f,y_f,t_f\mid x_0,y_0)\nonumber \\
&=&\left(\frac{\beta k}{2\pi}\right)\ \exp\Bigl[-\frac{\beta k \rme^{-2t_f/\tau_r}}{2b^2}\nonumber\\
&&\times \left( v^2\tau_r^2+ p\rme^{2 t_f/\tau_r}+2\rme^{t_f/\tau_r}v\tau_r q\right)\Bigr],\label{jointfinal}
\end{eqnarray}
where
\begin{eqnarray}
p&=&v^2t_f^2 b+2 \tau_r v x_f-2\sigma^2\tau_r^3 v x_f+(x_f^2+y_f^2)(1+\sigma^4\tau_r^4)\nonumber\\
&&+ \tau_r^2 [v^2+4\sigma v y_f+2\sigma^2(x_f^2+y_f^2)]-2t_f v x_f\nonumber\\
 &&-2 t_f v^2\tau_r- 2t_f v\tau_r \sigma [y_f+\sigma\tau_r(x_f+\sigma\tau_r y_f)],\\
 q&=&\cos(\sigma t_f)[vt_f-x_f+\tau_r (-v+\sigma^2\tau_rx_f-2\sigma y_f)]\nonumber\\
 &&+ \sin(\sigma t_f)[-y_f+\sigma\tau_r (-v t_f+2x_f+\sigma\tau_r y_f)].
\end{eqnarray}
During simplifications, we have used the relation,  
$\beta D k\tau_r=1$. For a particular set of parameter values, the joint probability density $\rho(x_f,y_f,t_f)$ is plotted in Fig.(\ref{fig:samplefig1}).

The marginal probability is 
defined as the probability  
of finding the particle at a given 
$x_f$ or $y_f$  for arbitrary values of $y_f$ or $x_f$, respectively. After integrating $\rho(x_f,y_f,t_f)$ with 
respect to the appropriate end point variable,  
the marginal probability distributions are 
found as 
\begin{widetext}
\begin{eqnarray}
\rho(x_f,t_f)&=&\int \rmd y_f\ \rho(x_f,y_f,t_f)\nonumber\\
&=&\left(\frac{k\beta}{2\pi}\right)^{1/2} \exp\biggl\{-\frac{k\beta \ \rme^{-2t_f/\tau_r}}{2b^4}\nonumber\\
&&\times \left(-v\ \rme^{t_f/\tau_r}[t_f b-\tau_r+\sigma^2\tau_r^3] +\rme^{t_f/\tau_r}x_f b^2 +v\tau_r\left[(-1+\sigma^2\tau_r^2)\cos(\sigma t_f)+2\sigma\tau_r\sin(\sigma t_f) \right]\right)^2\biggr\} ,\label{marginalxf}
\end{eqnarray}
and
\begin{eqnarray}
\rho(y_f,t_f)&=&\int \rmd x_f\ \rho(x_f,y_f,t_f)\nonumber\\
&=&\left(\frac{k\beta}{2\pi}\right)^{1/2} \exp\biggl\{-\frac{k\beta \ \rme^{-2t_f/\tau_r}}{2b^4}\nonumber\\
&&\times \left(-v\sigma\tau_r \ \rme^{t_f/\tau_r}\left[t_f b-2\tau_r\right] +y_f\ \rme^{t_f/\tau_r}\ b^2+ v\tau_r\left[-2\sigma\tau_r\cos(\sigma t_f) +(-1+\sigma^2\tau_r^2)\sin(\sigma t_f)\right]\right)^2\biggr\} .\label{marginalyf}
\end{eqnarray}
\end{widetext}
Eqs. (\ref{jointfinal}), (\ref{marginalxf}) 
and (\ref{marginalyf}) are the main results of this section.

\section{Generalized Onsager-Machlup Lagrangian and stochastic thermodynamic quantities}\label{sec:fewrelations}
In this section, we establish an energy-balance relation 
and a second law type formulation
 valid for a single trajectory. We consider that, for every forward trajectory $[x(t),y(t)]$ specifying the system's evolution over time interval $[-\tau,\tau]$, there exists a time-reversed trajectory $[x^{\dagger}(t),y^{\dagger}(t)]$ over the same interval. The  probability measure of the forward path could be expressed in terms of the transition probability up to a normalization factor as,
\begin{flalign}
&P_F(x(\tau),y(\tau)\mid x(-\tau),y(-\tau))&\nonumber\\ & 
\sim \exp\left(-\int_{-\tau}^{\tau} 
{\cal L}(x,y,\dot{x},\dot{y},t) \, \rmd t\right) ,&
\end{flalign} 
with the generalized Onsager-Machlup Lagrangian ${\cal L}(x,y,\dot{x},\dot{y},t)$ as given in 
 Eq.(\ref{om_lagrangian}). Similarly, the 
transition probability corresponding to the 
time reversed trajectory is expressed as \cite{taniguchi,cohen}
\begin{flalign}
&P_R(x^{\dagger}(\tau),y^{\dagger}(\tau)\mid x^{\dagger}(-\tau),y^{\dagger}(-\tau))&  \nonumber\\ &\sim \exp\Biggl(
-\int_{-\tau}^{\tau} {\cal L}^{\dagger}(x^{\dagger},y^{\dagger},\dot{x}^{\dagger},\dot{y}^{\dagger},t) \, \rmd t\Biggr),&
\end{flalign}
where ${\cal L}^{\dagger}$ is obtained from  ${\cal L}$ under the time-reversal changes $v \rightarrow -v,\ x^{\dagger}(t)\rightarrow x(-t),\ \dot{x}^{\dagger}(t)\rightarrow -\dot{x}(-t),\ y^{\dagger}(t)\rightarrow y(-t)$ and  $\dot{y}^{\dagger}(t)\rightarrow -\dot{y}(-t) $.

\subsection{Entropy production in the medium}
The ratio of the transition probability for the forward trajectory to that of the reversed trajectory is given by
\begin{flalign}
&\frac{P_F(x(\tau),y(\tau)\mid x(-\tau),y(-\tau))}{P_R(x^{\dagger}(\tau),y^{\dagger}(\tau)\mid x^{\dagger}(-\tau),y^{\dagger}(-\tau))}&\nonumber\\&= \exp\left[ -\int_{-\tau}^{\tau}({\cal L}-{\cal L}^{\dagger})\, \rmd t\right],& \label{ratio_trans}
\end{flalign}
where
\begin{equation}
{\cal L}-{\cal L}^{\dagger}=\beta \left[k \dot{x}(x-vt)+k\dot{y}y +n\dot{x}y-n\dot{y}x\right].\label{entropy}
\end{equation}
The entropy production in the medium $\Delta S_m$ satisfies the detailed fluctuation theorem \cite{park1} 
\begin{equation}
\Delta S_m = k_B\ \ln \left(\frac{P_F(x(\tau),y(\tau)\mid x(-\tau),y(-\tau))}{P_R(x^{\dagger}(\tau),y^{\dagger}(\tau)\mid x^{\dagger}(-\tau),y^{\dagger}(-\tau))}\right),\label{entropy1}
\end{equation} 
 From relation (\ref{ratio_trans}) and (\ref{entropy}), we have
\begin{flalign}
\Delta S_m=\frac{1}{T}\Biggl[-\Delta U+ \int_{-\tau}^{\tau} \frac{\partial U}{\partial t} \rmd t 
+ \int_{-\tau}^{\tau} (-n\dot{x}y+n\dot{y}x) \rmd t\Biggr],\label{entropy2}
\end{flalign}
where $\Delta U=U(x(\tau),y(\tau))-U(x(-\tau),y(-\tau))$ is the change in the potential energy and the explicit form of the second term is 
\begin{eqnarray}
\int_{-\tau}^\tau\frac{\partial U}{\partial t} dt=-k v\int_{-\tau}^{\tau} (x-v t).
\end{eqnarray} 

\subsection{Work and dissipated heat}
Since the heat transferred to the medium is 
${\cal Q}=T\Delta S_m$, we use Eq.  (\ref{entropy2}) to find 
\begin{flalign}
&{\cal Q}= \left[-\Delta U+ \int_{-\tau}^{\tau} \frac{\partial U}{\partial t}\ \rmd t\ + \int_{-\tau}^{\tau} (-n\dot{x}y+n\dot{y}x)\ \rmd t\right]&\nonumber\\
&=-\Delta U+ {\cal W}_{\rm tr}+ {\cal W}_{\rm m}.& \label{energy_conserv}
\end{flalign} 
In the last equality, the dissipated heat 
is expressed in terms of the change in the 
potential energy and the work that consists of 
two parts, namely, the thermodynamic work,  
${\cal W}_{\rm tr}=
\int_{-\tau}^{\tau} \frac{\partial U}{\partial t}\ \rmd t$, 
which is the work done on the system by the harmonic trap \cite{sekimoto,mazonka,zon_cohen1,zon_cohen2} 
and  the mechanical work, ${\cal W}_{\rm m}= \int_{-\tau}^{\tau} 
(-n\dot{x}y+n\dot{y}x)\ \rmd t$. 
Similar partition of the total work 
into physically meaningful works 
can be found in  recent literatures \cite{pjop,ciliberto2,hanggi}. 
For further discussions on various definitions of 
work and their  behavior under  gauge 
transformation, readers are referred to \cite{hanggi}. 
Together with these definitions, 
Eq.(\ref{energy_conserv}) becomes 
the statement of the energy conservation 
principle valid on  each trajectory \cite{seifert,seifert1,sekimoto}.

Next, to show the second law type formulation for the medium entropy production in the framework of the Onsager-Machlup 
theory \cite{taniguchi,cohen}, 
we split the Lagrangian ${\cal L}$ 
into time reversal symmetric, 
${\cal L}^{\rm r}={\cal L}+{\cal L}^{\dagger}$, and anti-symmetric, ${\cal L}^{\rm ir}={\cal L}-{\cal L}^{\dagger}$,  components  and arrive at
\begin{eqnarray}
{\cal L}=\frac{1}{2}\left[ {\cal L}^{\rm r} + {\cal L}^{\rm ir}\right]\ = \frac{1}{2}\left[ {\cal L}^{\rm r}-\Delta \dot{S}_m/k_B\right]. \label{dissipation_fn}
\end{eqnarray}
Using   Eqs. (\ref{ratio_trans}) and 
(\ref{entropy1}), ${\cal L}^{\rm ir}$ 
 in the above equation has been replaced by the rate of entropy production 
$\Delta \dot{S}_m$.  
It is evident from Eq.(\ref{om_lagrangian}) that 
 the average path described by 
\begin{eqnarray}
\left\langle\dot{x}\right\rangle+(\left\langle x\right\rangle-vt)/\tau_r +\sigma \left\langle y\right\rangle =0, \label{avgpath1}\\
\left\langle \dot{y}\right\rangle+\left\langle y\right\rangle/\tau_r -\sigma \left\langle x\right\rangle=0, \label{avgpath2}
\end{eqnarray}
 corresponds to the minimum value 
of the integral $\int_0^{t_f} {\cal L}(x,y,\dot{x},\dot{y},t) \,\rmd t $.
Substituting (\ref{avgpath1}) and (\ref{avgpath2}) into the 
 relation 
\begin{equation}
{\cal L}^{\rm r}\left(\langle\dot{x}\rangle,\langle x\rangle, \langle \dot{y}\rangle, \langle y\rangle, t\right) = \Delta \dot{S}_m \left(\langle\dot{x}\rangle,\langle x\rangle, \langle \dot{y}\rangle,\langle y\rangle, t\right) /k_B,
\end{equation}
 we find
\begin{eqnarray}
{\cal L}^{\rm r}&=&\frac{1}{4D}\left[\langle\dot{x}\rangle-(\langle x\rangle-vt)/\tau_r -\sigma \langle y\rangle\right]^2\nonumber\\
&&+\frac{1}{4D} \left[-\langle\dot{y}\rangle +\langle y\rangle/\tau_r -\sigma \langle x\rangle\right]^2  > 0 ;\label{positivity}
\end{eqnarray}
Eq. (\ref{positivity}), ensuring 
$\Delta \dot{S}_m \left(\langle\cdots \rangle\right) > 0$, 
shows the consistency  with the  
second law of thermodynamics.

\section{Distribution of the work performed by the moving trap}\label{sec:workdist1}
In this section, we derive the distribution of the work done by the harmonic trap on the Brownian particle. The work done or the energy put into the system by the trapping potential over time 
$[0:t_f]$ is 
\begin{equation}
{\cal W}_{\rm tr}[x(t)] = -k v\int_0^{t_f} (x-v t)\, {\rm d} t .
\end{equation}
In order to obtain the distribution of this work, we need to perform functional averages  of this quantity
 over all possible paths $\left[x(t),y(t); 0\le t\le t_f \right]$ as well as integrals over the initial and final points \cite{farago,taniguchi,cohen} of the path. The probability measure of such a path is given by,
\begin{flalign}
{\cal P}\left[x(t),y(t); 0\le t\le t_f \right] \sim \exp\left(-\int_0^{t_f} {\cal L}(x,y,\dot{x},\dot{y},t)\, \rmd t\right) .
\end{flalign}
So, the probability distribution of the work performed by the trap, henceforth denoted as $W_{\rm tr}$, is 
\begin{flalign}
&P(W_{\rm tr}) = \left\langle \delta \left(W_{\rm tr}-\left\lbrace -k v\int_0^{t_f} (x-v t)\, {\rm d} t \right\rbrace\right)\right\rangle & \nonumber\\
&=\int \rmd x_0 \int \rmd y_0\ \rho(x_0,y_0)\int \rmd x_f\int \rmd y_f&\nonumber\\
&\times \int {\cal D}[x(t)]\int {\cal D}[y(t)]\ {\cal P}\left[x(t),y(t)\right]&\nonumber\\  &\times \delta \left(W_{\rm tr}-\left\lbrace -k v\int_0^{t_f} (x-v t)\, {\rm d} t \right\rbrace\right)& \nonumber\\
&= \frac{\beta}{2\pi}\int_{-\infty}^{\infty} \rmd \lambda \, \rme^ {\rmi \lambda \beta W_{\rm tr}}\ P(\rmi \lambda W_{\rm tr});& 
\label{wd_trap} 
\end{flalign}
where
\begin{flalign}
&P(\rmi \lambda W_{\rm tr})=\int \rmd x_0 \int \rmd y_0\ \rho(x_0,y_0)\int \rmd x_f\int \rmd y_f& \nonumber\\
&\times \int {\cal D}[x(t)]\int {\cal D}[y(t)]\ \exp\left(-\int_0^{t_f} {\cal L}_M(x,y,\dot{x},\dot{y},\lambda)\, \rmd t\right),& \label{ft_wd_trap}
\end{flalign}
with 
\begin{eqnarray}
{\cal L}_M(x,y,\dot{x},\dot{y},\lambda)= \frac{1}{4D}\Bigl\{ \left[\dot{x}+(x-vt)/\tau_r +\sigma y\right]^2\nonumber\\
 + \left[\dot{y}+y/\tau_r -\sigma x\right]^2-4\rmi \lambda v(x-vt)/\tau_r \Bigr\} \label{modlagrangian}.
\end{eqnarray}
In order to derive the distribution function at a finite time, we consider the system to have evolved from the 
initial equilibrium distribution,  $\rho(x_0,y_0)=\frac{\beta k}{2\pi}\ \exp\left(-\frac{\beta k}{2}[x_0^2+y_0^2]\right)$.

To evaluate the functional integrals in (\ref{ft_wd_trap}), we follow the variational approach  discussed earlier 
in section \ref{sec:probdist}.  
We seek  the most probable trajectory 
$(\tilde{x}(t)$, $\tilde{y}(t))$ 
which corresponds to the largest contribution 
to the path integral.  The extremization 
 condition, $\delta \int_0^{t_f} {\cal L}_M\, \rmd t=0$, 
leads to  two Euler-Lagrange equations 
\begin{flalign}
&\ddot{\tilde{x}}+2\sigma \dot{\tilde{y}}-(\sigma^2 +1/\tau_r^2)\tilde{x} -v(1-2\rmi \lambda)/\tau_r +v t/\tau_r^2 =0,& \label{elex_wd_trap}\\
&\ddot{\tilde{y}}-(\sigma^2+1/\tau_r^2)\tilde{y} -2\sigma\dot{\tilde{x}}+v\sigma t/\tau_r =0,& \label{eley_wd_trap}
\end{flalign}
which are supplemented with the boundary conditions, $\tilde{x}(t=0)=x_0,\ \tilde{x}(t=t_f)=x_f,\ \tilde{y}(t=0)=y_0$ 
and  $\tilde{y}(t=t_f)= y_f$.
After some simplifications, the two Euler-Lagrange equations can be expressed in terms of higher derivatives of a single variable only. These are 
\begin{flalign}
&\ddddot{\tilde{y}}+2\left(\sigma^2-1/\tau_r^2\right)\ddot{\tilde{y}}+\left(\sigma^2 +1/\tau_r^2\right)^2\tilde{y}&\nonumber \\
&-\left(\sigma^2+1/\tau_r^2\right)v\sigma t/\tau_r +2\sigma v/\tau_r^2 =0\ ,& \\
\left(\sigma^2+1/\tau_r^2\right) \tilde{x} =& \frac{1}{2\sigma}\left[ \dddot{\tilde{y}}-(\sigma^2+1/\tau_r^2)\dot{\tilde{y}}+v\sigma/\tau_r \right]& \nonumber\\
&+2\sigma\dot{\tilde{y}} -v\bar{\lambda}/\tau_r +v t/\tau_r^2 \ ,&
\end{flalign} 
where, $\bar{\lambda} =1-2\rmi \lambda$. The optimal solutions are 
\begin{flalign}
\tilde{x}(t)=&v\left[bt-\bar{\lambda}\tau_r-(\bar{\lambda}-2)\sigma^2\tau_r^3\right]/b^2& \nonumber\\ 
&+\cosh(t/\tau_r)\left[c_2 \cos(\sigma t)-c_1\sin(\sigma t)\right]&\nonumber\\
&+ \sinh(t/\tau_r) \left[c_4 \cos(\sigma t)-c_3\sin(\sigma t)\right],&\label{mostprob_x}\\
\nonumber\\
\tilde{y}(t) = &\sigma \tau_r v [bt-2\tau_r]/b^2 + \cosh(t/\tau_r)[c_1 \cos(\sigma t)+c_2\sin(\sigma t)]&\nonumber\\
&+ \sinh(t/\tau_r)[c_3\cos(\sigma t)+c_4\sin(\sigma t)].&\label{mostprob_y}
\end{flalign}
The explicit dependence of the  
constants $c_1,\ c_2,\ c_3$ and $c_4$  on the boundary conditions and other parameters of the problem is shown in appendix \ref{app:appendixB}.
Incorporating the quadratic term arising from  the fluctuations around the most probable path, we write Eq.(\ref{ft_wd_trap}) 
as, \begin{flalign}
P(\rmi \lambda W_{\rm tr})&=\phi(t_f) \int \rmd x_0 \int \rmd y_0\ \rho(x_0,y_0)\int \rmd x_f&\nonumber\\
&\times\int \rmd y_f\ \exp\left(-\int_0^{t_f} {\cal L}_M(\tilde{x},\tilde{y},\dot{\tilde{x}},\dot{\tilde{y}},\lambda)\, \rmd t\right).&\label{ft_wd_trap3}
\end{flalign}
 Substituting the solutions (\ref{mostprob_x}) and (\ref{mostprob_y}) into (\ref{ft_wd_trap3}) 
and subsequently performing first the integral over time and then over   final and initial positions (some of the intermediate calculations are shown in appendix \ref{app:appendixB}), 
we have
\begin{eqnarray}
P(\rmi \lambda W_{\rm tr})= \exp\left[-\left(W_0 \lambda^2 +\rmi \bar{W}\lambda \right)\right], \label{ft_wd_trap2}
\end{eqnarray}
where 
\begin{eqnarray}
W_0 &=& \frac{v^2\rme^{-t_f/\tau_r}}{D b^2}\Bigl\{ \rme^{t_f/\tau_r}\left[bt_f-\tau_r+\sigma^2\tau_r^3\right]\nonumber\\
&&+\cos(\sigma t_f)\left[\tau_r-\sigma^2\tau_r^3 \right]-2\sigma\tau_r^2 \sin(\sigma t_f)\Bigr\} ;\label{w0}\\
\bar{W}&=&\frac{v^2 \rme^{-t_f/\tau_r}}{2Db^3} \Bigl\{ \rme^{t_f/\tau_r}\bigl[-2\tau_r+6\sigma^2\tau_r^3 +\sigma^2 b^2 t_f^2\tau_r \nonumber\\
&& +2t_f(1-\sigma^4\tau_r^4)\bigr]+\cos(\sigma t_f)\left[2\tau_r-6\sigma^2\tau_r^3\right] \nonumber\\
&&+2\sigma\tau_r^2 \sin(\sigma t_f)\left[\sigma^2\tau_r^2 -3\right]\Bigr\}. \label{wbar}
\end{eqnarray}
\begin{figure}[ht!]
  \centering
   \includegraphics[height=.36\textwidth]{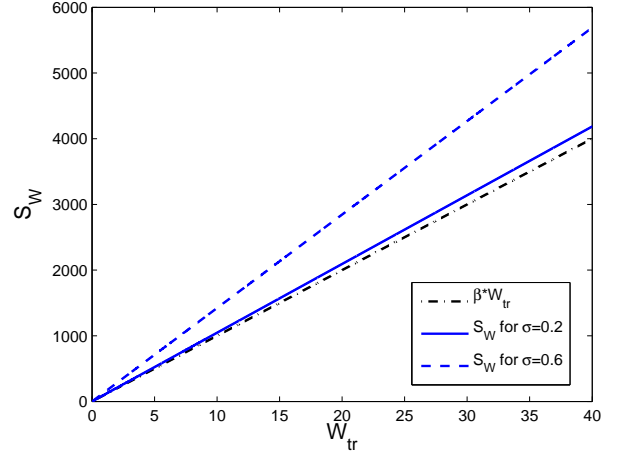}  
\caption{Plot of the symmetry function with $W_{\rm tr}$ for the parameter values $D=0.01,\  
\beta=100,$ $\tau_r=0.5,\ k=2,$ and $v=2$, at the final time $t_f=6$. The
 dashed-dot line corresponds to the  situation 
where the symmetry 
function follows the fluctuation theorem. 
The deviation from the fluctuation relation increases for larger values of $\sigma$.}
\label{fig:samplefig}
\end{figure}
Substituting (\ref{ft_wd_trap2}) into Eq.(\ref{wd_trap}) and doing the final Gaussian integral over $\lambda$, we find the final 
form of the work distribution as
\begin{eqnarray}
P(W_{\rm tr})=\left(\frac{\beta^2}{4\pi W_0}\right)^{1/2}\ \exp\left[-\frac{(\beta W_{\rm tr}-\bar{W})^2}{4W_0}\right].
\end{eqnarray}

 The known  result of the work distribution function for a uniformly dragged colloidal 
particle  \cite{mazonka,zon_cohen1,zon_cohen2,
seifert3,imparato3} can be 
retrieved by substituting $\sigma=0$ in our general result.  
The  work distribution function, in this case, obeys the fluctuation theorem. 
The symmetry function defined as 
$S_W=\ln\left[\frac{P(W_{\rm tr})}{P(-W_{\rm tr})}\right]$, therefore,  has  a  value 
$S_W=\beta W_{\rm tr}$.
In the presence of the non conserving force ($\sigma\neq 0$), 
the symmetry function is
\begin{equation}
S_W=\beta W_{\rm tr} \bar{W}/W_0 .
\end{equation}
 Clearly it does not satisfy the conventional 
fluctuation theorem since to hold this relation, it is required that $S_W=\beta W_{\rm tr}$. 
The present work distribution function satisfies the fluctuation theorem if $W_0=\bar{W}$. 
Given the following expressions for $W_0$ and $\bar{W}$
\begin{flalign}
&W_0 \approx  \frac{v^2}{Db^2}\left[ bt_f-\tau_r+\sigma^2\tau_r^3\right],&\\
&\bar{W} \approx  \frac{v^2}{2Db^3}\left[-2\tau_r + 6\sigma^2\tau_r^3+ \sigma^2b^2t_f^2\tau_r +2t_f(1-\sigma^4\tau_r^4)\right], &
\end{flalign} 
in the steady state limit {$t_f/\tau_r>>1$, such a 
condition is satisfied if 
\begin{eqnarray}
\sigma^2\tau_r \left[6\tau_r^2 -2\sigma^2\tau_r^4 +t_f^2 b^2-4t_f\tau_r b\right]=0. \label{condn}
\end{eqnarray}
 $\sigma=0$ trivially satisfies this equation. The other 
nonzero values of $\sigma$ for which Eq. (\ref{condn})  is 
satisfied are 
\begin{flalign}
&\sigma_1 =\frac{1}{t_f \tau_r}\left[2t_f\tau_r -t_f^2+\tau_r^2 -\tau_r^2 \sqrt{(-4(\frac{t_f}{\tau_r})^2+4\frac{t_f}{\tau_r}+1})\right]^{\frac{1}{2}}&\nonumber \\
&=-\sigma_2 ; & \\
&\sigma_3 =\frac{1}{t_f\tau_r}\left[2t_f\tau_r -t_f^2+\tau_r^2 + \tau_r^2 \sqrt{(-4(\frac{t_f}{\tau_r})^2+
4\frac{t_f}{\tau_r}+1})\right]^{\frac{1}{2}}&\nonumber \\
&=-\sigma_4. &
\end{flalign}
In the steady state limit  ($t_f/\tau_r>>1$), these roots 
become complex and hence not physically acceptable. 
To see the deviation from the fluctuation theorem, 
 we have plotted the symmetry function as a function of 
 $W_{\rm tr}$ for different values of $\sigma$ (see figure (\ref{fig:samplefig})). The deviation becomes more prominent 
as one increases the value of $\sigma$.

\section{Summary and perspectives}\label{sec:summary}
In this article, we have studied the two dimensional motion of a Brownian particle subject to a time dependent harmonic force and a nonconservative rotational torque. Starting with the overdamped Langevin equations, we have obtained a path integral description 
for  the transition probability of the Brownian particle. 
The transition probability gives the probability of 
finding the Brownian particle at a final  position at time
$t_f$  given that  it was at a specific position
 at an earlier time $t_0$. 
 Using a variational method, we evaluate the path integral to obtain 
the explicit form of the transition probability. This, upon integration over the initial position coordinates gives the 
joint probability distribution.

 In order to obtain the definitions of the work, 
the dissipated heat and the entropy production 
of the medium, we consider the  forward and time-reversed trajectories of the particle. 
Here, the time reversed trajectory is characterized  by the reversal of the velocity variables and the flipping of the end points with the  same underlying dynamics. After identifying the medium entropy production and the dissipated heat, we obtain  the energy conservation relation and, subsequently, define the two works; one done by the moving trap and the other by the 
 nonconservative force. Further, we show that the average path corresponding to the minimum of the generalized Lagrangian 
leads to the positive entropy production rate.

 The expression of the thermodynamic work  further allows us to derive its distribution function using   a similar path integral approach adopted for the transition probability.
We expressed the distribution function in a 
functional integral form which was approximated by the maximum 
contribution coming from the optimal path and the fluctuations upto second order   about the optimal path. The distribution is Gaussian in nature. The explicit form of the distribution correctly produces the 
known result of the work distribution
 for a dragged Brownian particle 
in the absence of any other force 
($\sigma=0$). However, in the 
presence of the 
nonconservative force, its mean and 
variance have a non-trivial oscillatory 
dependence with $\sigma$ as the angular frequency.  
As a consequence,  the work distribution  
fails to  satisfy the conventional work fluctuation relation.

In this context, we discuss another version of the model where 
the thermodynamic work naturally appears as a ratio 
of the forward and reverse process and satisfies fluctuation relation. Here, the nonconservative force is applied at the position of the minimum of the moving harmonic trap. In the reference frame of the trap, the particle moving under the potential  $\phi(X,y)=k(X^2+y^2)/2$, 
satisfies the overdamped Langevin equations given by,
\begin{eqnarray}
\dot{X}=-X/\tau_r -v -\sigma y+ \xi_x/\alpha ,\\
\dot{y}=-y/\tau_r+\sigma X+ \xi_y/\alpha ,
\end{eqnarray}
with, $X=x-vt$. 

To obtain the thermodynamic work and its fluctuation 
relation, we consider the time reversed process  
with the  reversal of sign in appropriate 
terms of the above equations \cite{vladimir}. 
The corresponding Lagrangian, denoted by $L^R$, has the form  
\begin{flalign}
&L^R=\frac{1}{4D}\left[ (-\dot{X}+X/\tau_r-v-\sigma y)^2 
+(-\dot{y}+y/\tau_r+\sigma X)^2\right].&
\end{flalign}
It is straightforward to obtain the 
ratio of the forward and reverse transition 
probabilities as
\begin{equation}
\frac{P_F}{P_R}= \exp \left[ \beta \left(-\Delta \phi -kv\int X\, \rmd t\right)\right] .\label{pfsummary}
\end{equation}
The first term 
 in the argument of the exponential  
stands for the change in the potential 
energy and the second term  can be recognized as the thermal work done \cite{jarzynskirel}. 
If we consider the initial distribution to be a canonical one,
i.e.  $\rho_0 \propto e^{-\beta \phi(X_0,y_0)}$, then the ratio of forward and 
reverse probability densities yields the fluctuation relation for the work ${\cal W}_{\rm tr}=\int -k v X \, \rmd t$,
\begin{equation}
\frac{{\cal P}_F}{{\cal P}_R}=\exp\left( \beta {\cal W_{\rm tr}}\right),
\label{rhofsummary}
\end{equation}
where the probability measure of the path  
starting at a given 
initial point is obtained from the transition probability as ${\cal P}_{F,R}=\rho_0\ P_{F,R}$. In this context, it will be of interest to verify whether the distribution of this work satisfies  various aspects of fluctuation relations and will be discussed elsewhere.

 Finally, the model studied in this paper and its extension seem to be promising for further investigations and the results obtained would also be amenable to experimental verifications.

\newpage
\appendix
\begin{widetext}
\section{Path integral solution of the Fokker-Planck equation}\label{app:appendixA}
The joint probability density function $\rho(x,y,t+\varepsilon)$ at time $t+\varepsilon$ is related to the transition probability\\ $P(x,y,t+\varepsilon\mid x^{\prime},y^{\prime},t)$ as
\begin{equation}
\rho(x,y,t+\varepsilon)=\int P(x,y,t+\varepsilon\mid x^{\prime},y^{\prime},t)\ \rho(x^{\prime}, y^{\prime},t)\, \rmd x^{\prime}\, \rmd y^{\prime} .
\end{equation}

For infinitesimal time difference $\varepsilon$, the transition probability can be expressed as,
\begin{eqnarray}
P(x,y,t+\varepsilon\mid x^{\prime},y^{\prime},t)\approx \left[ 1+\hat{L}_{\rm FP}\cdot \varepsilon +{\it O}(\varepsilon^2)\right]\delta(x-x^\prime)\ \delta(y-y^\prime) .
\end{eqnarray}

Using the Fourier representation of the $\delta$ function and substituting $\hat{L}_{\rm FP}$ from Eq.(\ref{lfp}), we have,
\begin{eqnarray}
P(x,y,t+\varepsilon\mid x^{\prime},y^{\prime},t)=\frac{1}{(2\pi)^2}\int_{-\infty}^{\infty}\rmd q_1 \int_{-\infty}^{\infty}\rmd q_2\biggl\{1+\varepsilon((x^\prime- vt)/\tau_r +\sigma y^\prime)\frac{\partial}{\partial x} +\varepsilon D\frac{\partial^2}{\partial x^2}\nonumber \\
-\varepsilon(-y^\prime /\tau_r +\sigma x^\prime)\frac{\partial}{\partial y}+\varepsilon D\frac{\partial^2}{\partial y^2}\biggr\}\ \exp\left[\rmi \left(q_1(x-x^\prime)+q_2(y-y^\prime)\right)\right]\nonumber\\
=\frac{1}{4\pi D\varepsilon}\ \exp \Biggl\{-\frac{1}{4D}\left(\left[\frac{(x-x^\prime)}{\varepsilon} +(x^\prime -vt)/\tau_r +\sigma y^\prime\right]^2\varepsilon +\left[\frac{(y-y^\prime)}{\varepsilon}+y^\prime/\tau_r -\sigma x^\prime\right]^2\varepsilon\right)\Biggr\} .
\end{eqnarray}

In order to obtain the transition probability for a finite time difference $(t_f-t_0)$, we divide the time difference into $N$ small segments of duration $\varepsilon = (t_f-t_0)/N$ . Then, in the limit $N\rightarrow \infty$ and $\varepsilon \rightarrow 0$, the repetitive application of the 
Chapman-Kolmogorov equation \cite{risken} yields, 

\begin{eqnarray}
P(x_f,y_f,t_f\mid x_0,y_0,t_0)&=&\int \prod_{i=1}^{N}\frac{\rmd x_i}{(4\pi D\varepsilon)^{1/2}} \int \prod_{i=1}^{N}\frac{\rmd y_i}{(4\pi D\varepsilon)^{1/2}}\nonumber \\
&& \times \exp\biggl\{-\sum_{i=1}^{N} \frac{1}{4D}\left( \left[(x_i-x_{i-1})/\varepsilon +(x_i -vt_i)/\tau_r+\sigma y_i\right]^2 +\left[(y_i-y_{i-1})/\varepsilon +y_i/\tau_r-\sigma x_i\right]^2\right)\, \varepsilon \biggr\}\nonumber\\
&=&\int {\cal D}[x(t)] \int {\cal D}[y(t)]\,\exp\biggl\{-\int_{t_0}^{t_f} \frac{1}{4D}\left( \left[\dot{x}+(x-vt)/\tau_r+\sigma y\right]^2 +\left[\dot{y}+y/\tau_r-\sigma x\right]^2\right)\, \rmd t\biggr\} .\nonumber\\
&&
\end{eqnarray}

\section{Some intermediate calculations of section \ref{sec:workdist1}} \label{app:appendixB}
After substituting the optimal solutions (\ref{mostprob_x}) and (\ref{mostprob_y}) into Eq.(\ref{modlagrangian}), the expression of the modified Lagrangian reads  
\begin{eqnarray}
{\cal L}_M(\tilde{x},\tilde{y},\dot{\tilde{x}},\dot{\tilde{y}},\lambda)&=&\frac{1}{4D\tau_r^2b^2}\bigl\{(-1+\bar{\lambda})v\tau_r -\rme^{t/\tau_r}b\left[(c_2+c_4)\cos(\sigma t)-(c_1+c_3)\sin(\sigma t)\right]\bigr\}^2\nonumber\\
&+&\frac{1}{4D\tau_r^2b^2} \bigl\{(-1+\bar{\lambda})\sigma\tau_r^2v +\rme^{t/\tau_r}b \left[(c_1+c_3)\cos(\sigma t)+(c_2+c_4)\sin(\sigma t)\right]\bigr\}^2\nonumber\\
&+&\frac{(\bar{\lambda}-1)v}{2D\tau_r}\biggl\{-\tau_r v[\bar{\lambda}b+\sigma^2 t\tau_rb-2\sigma^2 \tau_r^2]/b^2 +\cosh(t/\tau_r)[c_2\cos(\sigma t)-c_1\sin(\sigma t)]\nonumber\\&&+\sinh(t/\tau_r)[c_4\cos(\sigma t)-c_3\sin(\sigma t)]\biggr\}.
\end{eqnarray}

where, $c_1,\ c_2,\ c_3$ and $c_4$ have the form 
\begin{flalign}
&c_1=y_0+2\sigma\tau_r^2 v/b^2,& \\
&c_2=x_0+v\tau_r(\bar{\lambda}b-2\sigma^2\tau_r^2)/b^2,&\\
&c_3=\frac{1}{b^2}\biggl\{-\coth(t_f/\tau_r)[y_0b^2+2\sigma\tau_r^2v]+{\rm csch}(t_f/\tau_r)\bigl[-\sigma\tau_r v(t_f b-2\tau_r)\cos(\sigma t_f)+y_f b^2\cos(\sigma t_f)&\nonumber\\
&+v\sin(\sigma t_f)(t_f b-\bar{\lambda}\tau_rb+2\sigma^2\tau_r^3)-x_f b^2\sin(\sigma t_f)\bigr]\biggr\},&\\
&c_4=\frac{1}{b^2}\biggl\{-\coth(t_f/\tau_r)[v\tau_r(\bar{\lambda}b-2\sigma^2\tau_r^2)+x_0b^2] +{\rm csch}(t_f/\tau_r) \bigl[ v\tau_r(\bar{\lambda}b -2\sigma^2\tau_r^2)\cos(\sigma t_f)+(x_f b^2-v t_f b)\cos(\sigma t_f)&\nonumber\\
&+y_f b^2\sin(\sigma t_f)-\sigma \tau_r v(t_f b-2\tau_r)\sin(\sigma t_f)\bigr]\biggl\}.& 
\end{flalign}

We use the above expressions in 
Eq.(\ref{ft_wd_trap3}) to evaluate various integrations. After performing the time integration and integrations over 
the final positions, we have
\begin{eqnarray}
P(\rmi \lambda W_{\rm tr})&=&\int \rmd x_0\int \rmd y_0\ \rho(x_0,y_0)\ \exp\bigl\{-v \rme^{-t_f/\tau_r}(1-\bar{\lambda})[h_1 x_0+h_2 y_0+h_3]/(4Db^3)\bigr\}\nonumber\\
&=&\frac{\beta k}{2\pi}\exp\left(-v\rme^{-t_f/\tau_r}(1-\bar{\lambda})h_3/(4Db^3)\right)\int \rmd x_0 \exp\biggl\{-\left[\beta k x_0^2/2+v\rme^{-t_f/\tau_r}(1-\bar{\lambda})h_1x_0/(4Db^3)\right]\biggr\}\nonumber\\
&&\times \int \rmd y_0 \exp\biggl\{-\left[\beta k y_0^2/2+v\rme^{-t_f/\tau_r} (1-\bar{\lambda}) h_2 y_0/(4Db^3)\right]\biggr\}, \label{appnb}
\end{eqnarray}
where $h_1,\ h_2$ and $h_3$ are 
\begin{flalign}
&h_1=-2b^2\left[ \cosh(t_f/\tau_r)-\cos(\sigma t_f)+\sigma\tau_r\sin(\sigma t_f)+\sinh(t_f/\tau_r)\right], &\\
&h_2=-2b^2\left[\sigma\tau_r\cos(\sigma t_f)-\sigma \tau_r\rme^{t_f/\tau_r}+\sin(\sigma t_f)\right],\ {\rm and}&
\end{flalign}
\begin{flalign}
&h_3=2\tau_r v\cos(\sigma t_f)[\bar{\lambda}b-4\sigma^2\tau_r^2] +v\cosh(t_f/\tau_r)\left[8\sigma^2\tau_r^3-2\bar{\lambda}\tau_r b+\sigma^2 t_f^2\tau_r b^2+ bt_f(1+b\bar{\lambda}-3\sigma^2\tau_r^2)\right]&\nonumber\\
&-2\sigma \tau_r^2 v\sin(\sigma t_f)[2+b\bar{\lambda}-2\sigma^2\tau_r^2]+ v\sinh(t_f/\tau_r)\bigl[6\sigma^2\tau_r^3-(1+\bar{\lambda})\tau_r+ (\bar{\lambda}-1)\sigma^4\tau_r^5 &\nonumber\\
 &+\sigma^2 t_f^2\tau_r b^2+ t_f b(1+\bar{\lambda}b-3\sigma^2\tau_r^2 )\bigr].&
\end{flalign}
After doing the Gaussian integrals in Eq.(\ref{appnb}) and using the relation $1/(D^2\beta k)=\tau_r/D$, we arrive at Eq.(\ref{ft_wd_trap2}) in the main text.
\end{widetext}

\end{document}